\documentclass[aps,prl,twocolumn,superscriptaddress,showpacs,floatfix]{revtex4-2}
\usepackage{graphicx,bm,epsfig,textcomp,color,dcolumn,setspace,array,mathrsfs,amsmath,amssymb,gensymb,booktabs,url,hyperref,bibentry}
\begin{document}

\title{Superconducting triode effect in a quantum-dot Josephson junction with a biased top gate}
\author{Yu-Hang Li}
\affiliation{Department of Physics, Nankai University, Tianjin 300071, China}
\affiliation{State Key Laboratory of Surface Physics, Fudan University, Shanghai 200433, China}
\author{Xiaan Du}
\affiliation{International Center for Quantum Materials, School of Physics, Peking University, Beijing 100871, China}
\author{Hua Jiang}
\thanks{jianghuaphy@fudan.edu.cn}
\affiliation{Interdisciplinary Center for Theoretical Physics and Information Sciences, Fudan University, Shanghai 200433, China}
\affiliation{State Key Laboratory of Surface Physics, Fudan University, Shanghai 200433, China}
\author{X. C. Xie}
\affiliation{Interdisciplinary Center for Theoretical Physics and Information Sciences, Fudan University, Shanghai 200433, China}
\affiliation{International Center for Quantum Materials, School of Physics, Peking University, Beijing 100871, China}
\affiliation{Hefei National Laboratory, Hefei 230088, China}

\begin{abstract}
Non-reciprocal supercurrents enable non-dissipative rectification, holding great promise for superconducting electronics. Conventionally, this non-reciprocity, termed the superconducting diode effect, requires the simultaneous breaking of time-reversal and parity symmetries. Here, we propose a superconducting triode effect in an asymmetric quantum-dot Josephson junction coupled to an additional metallic top gate, which breaks the parity symmetry while explicitly preserving time-reversal symmetry. We demonstrate that the supercurrent across this junction exhibits a strong non-reciprocal effect that can be continuously manipulated via the top gate to achieve an ideal unidirectional supercurrent, thus manifesting a superconducting triode effect. Furthermore, under radio-frequency radiation, this junction exhibits highly asymmetric Shapiro steps, realizing fully quantized supercurrent rectification. Our work not only provides an alternative physical mechanism for the superconducting diode effect observed in Josephson junctions with explicit time-reversal symmetry, but also introduces a new tuning knob to manipulate supercurrent non-reciprocity.
\end{abstract}

\maketitle

The superconducting diode effect (SDE)~\cite{ando_observation_2020,Margarita2022,Daido2022,lyu_superconducting_2021,wu_field-free_2022,narita_field-free_2022,jeon_zero-field_2022}, allowing the supercurrent to flow preferentially in one direction, has recently attracted significant interest for its great promise in non-dissipative circuits~\cite{Nadeem2023The,Jiang2022Superconducting,Ma2025Superconducting,Zhang2022General,Shaffer2025theories,moll_evolution_2023}. To date, the SDE has been realized in various systems including topological Josephson junction~\cite{baumgartner_supercurrent_2022,pal2022josephson,sundaresh2023diamagnetic,qin2017superconductivity,Yasuda2019Nonreciprocal,Cui2019Transport}, single atomic Josephson junction~\cite{trahms2023diode}, twisted materials~\cite{diez2023symmetry,de_vries_gate-defined_2021,lin_zero-field_2022}, and bulk single-phase superconductors~\cite{le_superconducting_2024,Yuki2020Non,Hou2023Ubi,gutfreund_direct_2023}. Fundamentally, because parity ($\mathcal{P}$) and time-reversal ($\mathcal{T}$) symmetries restrict the current-phase relation to be exactly antisymmetric, i.e., $I(\phi)=-I(-\phi)$, the emergence of this SDE in principle requires the simultaneous breaking of both parity and time-reversal symmetries~\cite{Margarita2022,Daido2022,He_2022,Yuan2022Super,wang2022symmetry}. However, recent experiments have reported the SDE in nominally $\mathcal{T}$-invariant system~\cite{golod_demonstration_2022,qi_high-temperature_2025,li2025field}, posing a profound challenge to this conventional symmetry paradigm. While it has been proposed that spontaneous $\mathcal{T}$ symmetry breaking in these experiments could possibly arise from trapped vortices induced by minute residual magnetic fields exceeding the lower critical field~\cite{golod_demonstration_2022}, a comprehensive understanding of the SDE in the absence of explicit $\mathcal{T}$ symmetry breaking is still under intense debate.

Beyond the dc regime, under a combined dc bias and radio-frequency (rf) radiation, recent studies reveal that these systems can exhibit asymmetric Shapiro steps~\cite{lou_quantized_2026,wang_quantum_2026}. In this dynamic regime, the time-averaged voltage across the Josephson junction vanishes for one current direction, but becomes strictly quantized in units of the rf frequency along the opposite direction, thereby enabling quantized supercurrent rectification. The underlying physical mechanism driving these asymmetric Shapiro steps, particularly in the absence of explicit $\mathcal{T}$ symmetry breaking~\cite{wang_quantum_2026}, also remains a subject of intense investigation. Furthermore, practical device integration demands an ideal SDE where the supercurrent flows exclusively in one direction while being prohibited completely in reversed direction~\cite{Devoret2013}. Nevertheless, the SDE is currently observed only as a partial asymmetry between the forward and reverse critical currents within a narrow parameter window, regardless of the degree of $\mathcal{P}$ and $\mathcal{T}$ symmetry breaking. Consequently, achieving a robust and ideal SDE poses another highly sought-after yet elusive goal.

\begin{figure}[ttt]
  \centering
  \includegraphics[width=0.98\linewidth]{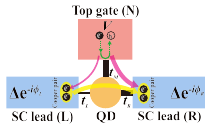}
  \caption{
	Schematic for the STE in an asymmetric quantum-dot Josephson junction with an additional biased top gate (N). Here, the coupling strength between the quantum-dot and the three leads are $t_L$, $t_R$, and $t_M$, respectively. The green line in top gate illustrates the Andreev reflection under a bias $V$. The yellow arrow illustrates the tunneling supercurrent between left (L) and right (R) superconducting leads driven by the phase difference $\phi=\phi_L-\phi_R$. The cyan arrows highlight the normal-metal-assisted injection currents, mediated by non-equilibrium Andreev processes involving the top gate (N).
	  	}
\label{fig1}
\end{figure}

In this Letter, we propose a superconducting triode effect (STE) in a hybrid three-terminal junction with a quantum-dot centered between two superconducting leads and a metallic top gate (Fig.~\ref{fig1}), which preserves $\mathcal{T}$ symmetry explicitly whereas breaks $\mathcal{P}$ symmetry if the coupling strengths between the quantum dot and the superconducting leads are asymmetric.  In particular, a small bias voltage on top gate can trigger a non-equilibrium Andreev supercurrent proportional to applied voltage, which is thus intrinsically immune to the constrain of the time-reversal symmetry. Remarkably, this phase coherent Andreev supercurrent asymmetrically shifts the current phase relation (CPR) between two superconducting leads, resulting in a highly tunable non-reciprocal supercurrent. Under optimal gate tuning, this shift can render the critical current entirely unidirectional, achieving the elusive ideal SDE. Furthermore, under a dc bias across the superconducting leads combined with rf radiation, the system naturally exhibits asymmetric Shapiro steps when the dc bias is commensurate with the radiation, enabling quantized supercurrent rectification. Since the non-reciprocal efficiency is exquisitely sensitive to the top gate bias, our theory thus introduces a powerful yet purely electrical tuning knob for manipulating supercurrent non-reciprocity.

The hybrid triode junction illustrated in Fig.~\ref{fig1} can be described by the Hamiltonian 
\begin{align}
\mathcal{H}=H_C+H_L+H_R+H_M+H_T. 
\label{sys_Ham}
\end{align}
Here, $H_C=\epsilon_0\sum_{\sigma}c_{\sigma}^{\dagger}c_{\sigma}$ is the Hamiltonian of the central quantum-dot, where $\epsilon_0$ is the onsite energy, and $c_{\sigma}^{\dagger}$ ($c_{\sigma}$) is the creation (annihilation) operator for a state with spin $\sigma$. $H_{\alpha=L,R,M}=\sum_{k\sigma}\epsilon_{\alpha k\sigma}a_{\alpha k\sigma}^{\dagger}a_{\alpha k\sigma}+\sum_{k}(\Delta_{\alpha}a_{\alpha k\uparrow}^{\dagger}a_{\alpha -k\downarrow}^{\dagger}+\textrm{H.c.})$ is the Hamiltonian for the $s$-wave superconducting/metallic lead, with the first term being the kinetic energy while the second term the superconducting pairing. In particular, the superconducting order parameter $\Delta_{L,R}=\Delta_0 e^{-i\phi_{L,R}}$ with $\Delta_0$ the uniform superconducting gap and $\phi_{L,R}$ corresponding phase, while $\Delta_M=0$ for the metallic top gate. $H_T=\sum_{\alpha k\sigma}(t_{\alpha}a_{\alpha k\sigma}^{\dagger}c_{\sigma}+\textrm{H.c.})$ describes the coupling between the central quantum-dot and all the leads, where $t_\alpha$ is the spin-conserving tunneling amplitude and H.c. is the shorthand notation for Hermitian conjugate. Crucially, the full Hamiltonian in Eq.~\ref{sys_Ham} preserves $\mathcal{T}$ symmetry explicitly whereas intrinsically breaks $\mathcal{P}$ symmetry if the coupling is asymmetric with $t_L\ne t_R$. Throughout the work, we take the superconducting gap $\Delta_0=1$ as the energy unit.
The average current $I_{\alpha}(t)$ flowing from $\alpha$ lead into the central quantum-dot can be obtained by using the Keldysh Green's function method, which is detailed in Supplementary Materials~\cite{SM}. The net supercurrent across the two superconducting leads is given by $I_s(t)=[I_L(t)-I_R(t)]/2$. 

We first consider the purely dc case where the voltage is allowed exclusively to the third metallic top gate while the superconducting leads are grounded. In this case, the supercurrent can be rigorously calculated through the equation $I_{\alpha}=({2e}/{h})\int{\textrm{d}\epsilon}\textrm{Re}\textrm{Tr}\{\sigma_z[G^r(\epsilon)\Sigma_{\alpha}^{<}(\epsilon)+G^<(\epsilon)\Sigma_{\alpha}^{a}(\epsilon)]\}$~\cite{MartnRodero2011JosephsonAA,SM}, where $e$ is the electron charge, $h$ is the Planck constant, $\sigma_z$ is the Pauli matrix. $G^{r,<}(\epsilon)$ are the Green's functions defined on the Nambu space $\begin{pmatrix}c_{\uparrow}, c_{\downarrow}^{\dagger} \end{pmatrix}^{\textrm{T}}$, which can be obtained by using the Dyson equation $G^{r/a}(\epsilon)=[\epsilon\pm i\gamma-H_C-\sum_{\alpha}\Sigma_{\alpha}^{r/a}(\epsilon)]^{-1}$ and the Keldysh equation $G^{<}(\epsilon)=G^r(\epsilon)\sum_{\alpha}\Sigma_{\alpha}^{<}(\epsilon)G^a(\epsilon)$. The terms $\Sigma^{r,a,<}_{\alpha}(\epsilon)$ represents the self-energies arising from the coupling to lead $\alpha$, satisfying the Hermitian relation $\Sigma^{r}_{\alpha}(\epsilon)=[\Sigma^a_{\alpha}(\epsilon)]^\dagger$. Besides,  adhering to the fluctuation-dissipation theorem for the uncoupled free leads, the distribution self energy is given by $\Sigma_\alpha^{<}(\epsilon)=f_{\alpha}(\epsilon)[\Sigma^a_{\alpha}(\epsilon)-\Sigma^r_{\alpha}(\epsilon)]$, with $f_{\alpha}(\epsilon)$ the diagonal matrix of the Fermi-Dirac distribution functions for electrons and holes. Re and Tr denote the real part and the trace over the Nambu space, respectively.

\begin{figure}[ttt]
  \centering
  \includegraphics[width=0.98\linewidth]{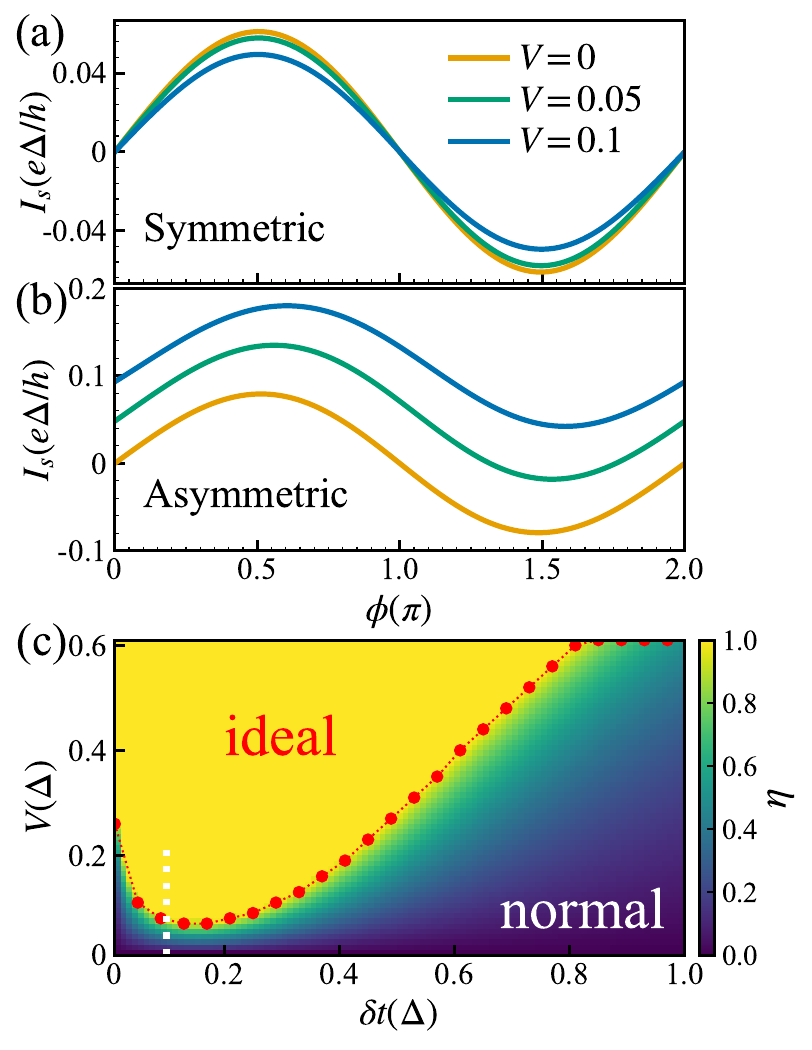}
  \caption{
	(a) and (b) Current-phase relations $I_s(\phi)$ for a symmetric junction with $t_L=t_R=0.1$ (a) and an asymmetric junction with $t_L=0.2$, $t_R=0.1$ (b) under different top gate bias $V$. Here, $t_M=0.5$.
	(c) Non-reciprocal supercurrent efficiency $\eta$ on the $V$-$\delta t$ plane. The red dashed line highlights the critical phase boundary with $\eta=1$, which separates the normal non-reciprocal regime and the ideal regime that only allows the unidirectional supercurrent. The white dashed line highlights the cases in (a) and (b). Here, $\delta t=t_L-t_R$, $t_R=0.1$ and $t_M=0.5$.
	  	}
\label{fig2}
\end{figure}

Figures~\ref{fig2}(a) and (b) show the CPR $I_s(\phi)=[I_L(\phi)-I_R(\phi)]/2$ under various top gate voltages $V$ for a symmetric junction with $t_L=t_R$ (a) and an asymmetric junction with $t_L\ne t_R$ (b), where $\phi=\phi_L-\phi_R$ denotes the macroscopic superconducting phase difference. For the symmetric junction [Fig.~\ref{fig2}(a)], the CPR follows a typical sinusoidal profile with strictly equal critical supercurrent magnitudes along positive and negative directions, in agreement with a short weak-link Josephson junction where the supercurrent non-reciprocal effect is absent because of the inherent $\mathcal{P}$ symmetry~\cite{Golubov2004The}. Here, the critical supercurrent amplitude  $I_c$ monotonically decreases as the increase of the top gate $V$, which can be ascribed to the decoherence effect from the particle leakage to the metallic top gate~\cite{MartnRodero2011JosephsonAA,pillet_andreev_2010}. In stark contrast, for the asymmetric junction in Fig.~\ref{fig2}(b), the CPR is deterministically shifted vertically by applied top gate voltage, rendering a controllable non-reciprocity effect. Remarkably, at $eV=0.1\Delta_0$, the CPR becomes fully positive as indicated by the brilliant blue line in Fig.~\ref{fig2}(b), signifying an ideal non-reciprocity effect where the supercurrent is only allowed to flow in one direction while blocked completely in the other direction. Consequently, our theory introduces a STE, establishing the third-terminal top gate as a robust tuning knob for controlling macroscopic supercurrent non-reciprocity. 

To elucidate the underlying physical origin of this STE, we explore the supercurrent in each lead separately. In principle, the supercurrent flowing from the superconducting lead $\alpha$ can be expressed in terms of the terminal-resolved components $I_{\alpha}=\sum_{\beta}I_{\alpha}^{\beta}$, where $I_{\alpha}^{\beta}$ represents specific current contribution flowing from lead $\alpha$ to lead $\beta$~\cite{SM}. For the superconducting lead $\alpha=L,R$, these terminal-resolved components originate entirely from coherent tunneling supercurrent given by $I_{\alpha}^{\beta}=({2e/h})\int{\textrm{d}\epsilon}\textrm{ReTr}[\sigma_z(G^r\Sigma_{\beta}^<G^a\Sigma_{\alpha}^a+\Sigma_{\beta}^rG^r\Sigma_{\alpha}^<G^a)]$. On the contrary, in the limit of small top gate voltage $V\ll\Delta$, direct quantum tunneling from the top gate into the superconducting leads is exponentially suppressed. The supercurrent flowing from the top gate is instead dominated by Andreev reflections, which is given by $I_{M}^{M}=({e/h})\int{\textrm{d}\epsilon}\textrm{Tr}\{\sigma_z[G^r\Sigma_{M}^{>}G^a\Sigma_{M}^{<}+G^r\Sigma_{M}^{<}G^a\Sigma_{M}^{>}]\}$. Further imposing steady-state charge conservation condition of $I_L+I_R+I_M=0$ yields the following relations for the bipartite components
\begin{align}\label{current_equation}
\begin{split}
I_{L}^{R}+I_{R}^{L}=0,\quad I_{L}^{M}+I_R^M+I_M^M=0.
\end{split}
\end{align}
Moreover, since the self-energies scale as $\Sigma_{\alpha}\propto t_{\alpha}^2$, it ensures that each bipartite current component $I_{\alpha}^{\beta}$ is rigorously proportional to $t_{\alpha}^2t_{\beta}^2$. The terminal-resolved components from the superconducting lead to the top gate is thus $I_{\alpha=L,R}^M=-t_{\alpha}^2I_M^M/(t_{L}^2+t_R^2)$. Consequently, the net traversing supercurrent across superconducting leads, $I_s=[I_{L}-I_{R}]/2$, can be analytically decoupled into two distinct physical processes in the form of
\begin{align}\label{current_equation_two}
\begin{split}
I_{s}=\frac{I_L^R-I_R^L}{2}+\frac{(t_R^2-t_L^2)I_M^M}{2(t_L^2+t_R^2)}.
\end{split}
\end{align}
The first term in Eq.~\ref{current_equation_two} constitutes the direct coherent Cooper-pair tunneling, analogous to the supercurrent in a conventional two-terminal Josephson junction~\cite{MartnRodero2011JosephsonAA}, albeit suppressed by the decoherence induced by the metallic top gate. The second term, on the other hand, captures the non-equilibrium injection current driven by the metallic top gate, which is intrinsically mediated by Andreev reflections involving all three terminals and provides the fundamental source of the spatial non-reciprocity.

When $V=0$, both $I_{L}^{M}$ and $I_{R}^{M}$ are identically zero due to the absence of a net non-equilibrium Andreev injection. The top gate assisted supercurrent in Eq.~\ref{current_equation_two} vanishes, restoring a conventional CPR devoid of any non-reciprocal effect. Similarly, in a symmetric junction with $t_L=t_R$, $I_{L}^{M}$ and $I_{R}^{M}$ exactly cancel each other owing to the symmetric scaling $t_{L}^2t_M^2=t_{R}^2t_M^2$. In this scenario, the top gate only acts as a symmetric source of decoherence, suppressing the overall critical current without breaking macroscopic reciprocity as shown in Fig.~\ref{fig2}(a). Nevertheless, in an asymmetric junction under finite top gate bias, the second term in Eq.~\ref{current_equation_two} yields a robust, voltage-dependent non-equilibrium current, which fundamentaly shifts the baseline of the current-phase relation, synergizes with the first direct-tunneling component to break the inversion symmetry of the total CPR, and in turn leads to a non-reciprocal supercurrent. It is this precise non-equilibrium interplay that gives rise to the STE, empowering the top gate to reliably and continuously tune the supercurrent non-reciprocity.

To systematically quantify the gate-tunable non-reciprocity, we calculate the non-reciprocal supercurrent efficiency defined as $\eta=(I_c^++I_c^-)/(I_c^++|I_c^-|)$, where $I_c^{\pm}$ are the maximum and minimum extrema of the current-phase relation, respectively. In Fig.~\ref{fig2}(c), we display the efficiency $\eta$ on the $V-\delta t$ plane, where $\delta t=t_L-t_R$ characterizes the spatial asymmetry of the junction. Overall, the efficiency $\eta$ increases with top gate voltage $V$. Notably, figure~\ref{fig2}(c) reveals two different regimes separated by the red dashed line marking the exact onset of the ideal non-reciprocity. In the standard regime below this boundary, the supercurrent remains bidirectional where $I_c^{+}$ and $I_c^{-}$ flow along opposite directions. Nevertheless, in the ideal regime above this critical boundary, the entire current-phase relation is elevated above zero as illustrated by the brilliant blue line in Fig.~\ref{fig2}(b), yielding a saturated efficiency of $\eta=1$ and unequivocally confirming the realization of an ideal superconducting rectification.

\begin{figure*}[ttt]
  \centering
  \includegraphics[width=0.95\linewidth]{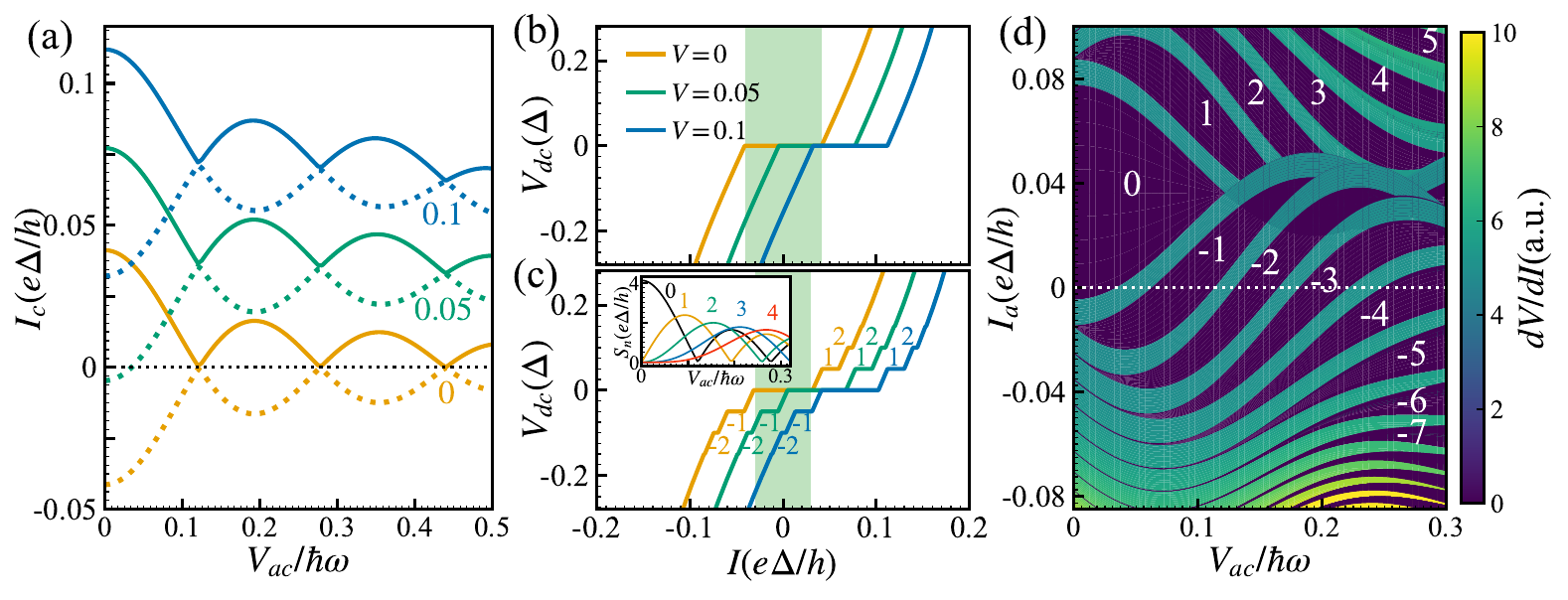}
  \caption{
	(a) Maximum $I_c^+$ (solid lines) and minimum $I_c^-$ (dashed lines) critical supercurrents of $I_s(\phi)$ as a function of the scaled radiation amplitude $eV_{ac}/\hbar\omega$ for different top gate bias $V$. 
	(b) and (c) $I$-$V$ curve for the junction under different top gates $V$ in the absence [(a), $V_{ac}=0$] and presence [(b), $eV_{ac}/\hbar\omega=0.05$] of radiation. The green shadings in the background highlight the symmetric supercurrent at $V=0$, serving as a benchmark. The inset displays five representative Shapiro step widths versus scaled radiation amplitude at $V=0$.
	(d) Differential resistance $dV/dI$ versus total time-averaging current $I_{a}$ and scaled radiation amplitude $eV_{ac}/\hbar\omega$ at a representative top bias $V=0.05$. The white dashed line marks the $I_a=0$ axis. The Shapiro step orders are labeled in the respective panels.
	Here, $t_L=0.2$, $t_R=0.1$, and $t_M=0.5$. The radiation frequency is $\hbar\omega=0.1$.
	  	}
\label{fig3}
\end{figure*}

We next turn to the ac supercurrent driven by a dc bias $V_{dc}$ cross the superconducting leads superimposed with a rf radiation characterized by an oscillating voltage $V_{ac}\cos{\omega t}$, where $V_{ac}$ and $\omega$ are the amplitude and frequency of the radiation. In this case, the onsite energy of each parts, including all leads and the central quantum dot, acquires an additional temporal-dependent modulation $\epsilon_{\alpha}(t)=\epsilon_{\alpha}-eV_{\alpha}^{dc}-eV_{\alpha}^{ac}\cos{\omega t}$~\cite{Cuevas1996,Cuevas2002}. Concurrently, the superconducting gap function dynamically evolves as $\Delta_{\alpha}=\Delta_0\textrm{exp}[-i\phi_{\alpha}-(2ie/\hbar)\int_0^t (V_{\alpha}^{dc}+V_{\alpha}^{ac}\cos{\omega\tau}) d\tau]$~\cite{Sun2000}. Using a gauge transformation, we can completely eliminate the explicit time dependence in the lead and central Hamiltonians, and transfer it entirely into the tunneling Hamiltonian $H_T(t)$ as a temporal dependent phase factors~\cite{Sun2000}. The system thus possesses two fundamental driving frequencies $\omega_1=2eV_{dc}/\hbar$ and $\omega_2=\omega$, and the supercurrent from each lead can then be expanded as a double Fourier series $I_{\alpha}(t)=\sum_{n,m}I_m^n\textrm{exp}[i(n\omega_1+m\omega_2)t]$~\cite{Cuevas2002,Li2018}. Here, the Fourier coefficients $I_m^n$ encode the multiphoton harmonic amplitudes with $I_0^0$ the time-averaged background current. When these two fundamental frequencies are commensurate, the system exhibits phase-locked Shapiro steps $\sum_{n,m}I_m^n\delta(2neV_{dc},m\hbar\omega)$~\cite{shapiro1963}, where $m$, $n$ are arbitrary integers, $\delta(2neV_{dc},m\hbar\omega)$ is the Kronecker delta function. In the absence of radiation with $V_{ac}=0$, the current returns to the DC supercurrent if $V_{dc}=0$, in which case $I_0^0=I_s$.

By virtue of the gauge invariance, we adopt a symmetric voltage configuration for the superconducting leads with $V_L(t)=-V_R(t)=(V_{dc}+V_{ac}\cos{\omega t})/2$ while taking $V_M(t)=V_C(t)=0$ for analytical convenience. Under this gauge, the time-averaged net dc supercurrent across the junction $I=(I_{L}-I_{R})/2$ can be decomposed as $I=I_s(\phi)+I_b(V_{dc})+\sum_{n,m}I_m^n\delta(2neV_{dc},m\hbar\omega)$. Here, the first term $I_s(\phi)$ corresponds to the 0th-order Shapiro step, the second term $I_b(V_{dc})$ accounts for the dissipative background current, and the third term encapsulates the higher-order Shapiro steps.

Figure~\ref{fig3}(a) displays the critical supercurrents $I_c^+$ and $I_c^-$ versus scaled radiation amplitude $eV_{ac}/\hbar\omega$ under different top gate voltages $V$ for an asymmetric junction with $t_L=0.1$, $t_R=0.2$. At $V=0$, $I_c^+$ and $I_c^-$ are perfectly symmetric with respect to zero, both of which evolves according to the standard Bessel-function modulation. In the presence of top gate voltage, $I_c^+$ and $I_c^-$ are uniformly elevated, indicating a robust vertical shift of the entire current-phase relation in agreement with the dc analysis in Fig.~\ref{fig2}(b). Consequently, the system can be dynamically driven into the unidirectional transport regime through the rf radiation, thereby demonstrating that the STE is robustly preserved even under strong rf radiation. Figure~\ref{fig3}(b) shows corresponding $I$-$V$ curve in the absence of radiation for the same asymmetric junction under different top gate bias $V$. This curve agrees with that in a typical Josephson junction~\cite{Golubov2004The}, where a dissipationless supercurrent branch strictly resides at zero bias voltage ($V_{dc}=0$). It smoothly transitions into an Ohmic behavior when the current exceeds the critical supercurrent. Besides, a finite top gate voltage $V$ induces a pronounced shift of the zero-voltage supercurrent window along the current axis., which originates from aforementioned normal-metal-assisted injection current, and is in general proportional to applied bias voltage. This shift underpins the fundamental physical mechanism for the emergence of asymmetric higher-order Shapiro steps when sweeping the applied dc current. As shown in Fig.~\ref{fig3}(c), in the jointly presence of both bias and radiation, the Shapiro steps become highly asymmetric under top gate voltage, demonstrating a quantized rectification where the voltage is zero along one direction while is quantized in units of the radiation along the reverse direction. Moreover, the evolution of the Shapiro steps versus the scaled radiation amplitude, plotted in inset of Fig.~\ref{fig3}(c) 
coincides quantitatively with expected Bessel functions. Given that the $0$-th and high-order Shapiro steps are obtained independently from different computational approaches~\cite{SM}, this unified convergence robustly validates the internal consistency and reliability of our theory.

To close our discussion, we finally calculate the differential resistance $dV/dI$ as a function of the total time-averaging current $I_a$ and scaled radiation amplitude $eV_{ac}/\hbar\omega$ for a representative top gate $V=0.05$, which is displayed in Fig.~\ref{fig3}(d). 
As the microwave driving amplitude increases, higher order Shapiro steps consecutively emerge. Strikingly, while the width evolution of these steps strictly follows the expected Bessel-function modulation, their distribution over the current polarity is highly asymmetric. We observe only five Shapiro steps in the positive current direction, compared to ten distinct steps in the reversed negative direction.
This stark asymmetry agrees quantitatively with recent experimental observations~\cite{lou_quantized_2026,wang_quantum_2026}, compellingly reaffirming a robust quantized supercurrent rectification

In summary, we have proposed a STE in an asymmetric quantum-dot Josephson junction with an additional biased top gate. We demonstrate that, in the purely dc case, this system exhibits a reliable supercurrent non-reciprocity, which can be manipulated in a large scale towards the elusive ideal case through the top gate voltage. Furthermore, under combined dc bias and rf radiation, the system exhibits highly asymmetric Shapiro steps, manifesting a striking quantized supercurrent rectification. Although our explicit calculations focused on specific coupling parameters in the weak-link limit, the underlying mechanism of this emergent STE relies on fundamental non-equilibrium Andreev processes. Consequently, our theory is universal and can be generalized to arbitrary coupling strengths, spanning from the weak-link to the strong-coupling regimes.

In practical experiments, the critical supercurrent is typically measured using a standard four-terminal geometry~\cite{golod_demonstration_2022,qi_high-temperature_2025,li2025field,lou_quantized_2026,wang_quantum_2026}, where a drive current is applied across the two outer leads while the voltage drop is detected between the two inner probes. Because the coupling strengths between these probes and the active junction region are inevitably asymmetric, our proposed triode mechanism could inadvertently emerge if parasitic current leakage occurs at the measurement probes. Therefore, our theory not only introduces a robust while purely electrical paradigm for manipulating macroscopic supercurrent non-reciprocity, but also provides a compelling alternative microscopic origin for the SDE recently observed in nominally $\mathcal{T}$-symmetry invariant systems.

This work is supported by the National Basic Research Program of China (Grants Nos. 2024YFA1409003 and 2022YFA1403700), National Natural Science Foundation of China (Grants Nos. 12147126 and 12404056). Yu-Hang Li also acknowledges the financial support from the Natural Science Foundation of Tianjin, China (Grant No. 24JCQNJC01910), the Fundamental Research Funds for the Central Universities (Grant No. 010-63263112), the State Key Laboratory of Surface Physics and the Department of Physics at Fudan University.

\bibliography{ref}

\end{document}


\title{Supplemental material for ``Superconducting triode effect in a hybrid quantum-dot junction"}

\author{Yu-Hang Li}
\affiliation{Department of Physics, Nankai University, Tianjin 300071, China}
\affiliation{State Key Laboratory of Surface Physics, Fudan University, Shanghai 200433, China}
\author{Xiaan Du}
\affiliation{International Center for Quantum Materials, School of Physics, Peking University, Beijing 100871, China}
\author{Hua Jiang}
\thanks{jianghuaphy@fudan.edu.cn}
\affiliation{Interdisciplinary Center for Theoretical Physics and Information Sciences, Fudan University, Shanghai 200433, China}
\affiliation{State Key Laboratory of Surface Physics, Fudan University, Shanghai 200433, China}
\author{X. C. Xie}
\affiliation{Interdisciplinary Center for Theoretical Physics and Information Sciences, Fudan University, Shanghai 200433, China}
\affiliation{International Center for Quantum Materials, School of Physics, Peking University, Beijing 100871, China}
\affiliation{Hefei National Laboratory, Hefei 230088, China}

\date{\today}

\maketitle

As discussed in the main text, the supercurrent flowing from lead $\alpha$ is calculated using the Keldysh non-equilibrium Green's function (NEGF) formalism. In this framework, the current can be expressed as
\begin{align}\label{current_def}
\begin{split}
I_{\alpha}(t)&=e\langle \partial_t \hat{n}_{\alpha}(t)\rangle=\frac{ie}{\hbar}\langle[\hat{n}_{\alpha}(t),\ H]\rangle\\
&=\frac{ie}{\hbar}\sum_{k\sigma}\left[t_{\alpha}(t)\langle a_{\alpha k\sigma}^{\dagger}(t)c_{\sigma}(t)\rangle-t_{\alpha}^*(t)\langle c_{\sigma}^{\dagger}(t)a_{\alpha k\sigma}(t)\rangle\right],
\end{split}
\end{align}
where $e$ is the elementary charge, $\hbar$ is the reduced Planck constant, and $\hat{n}_{\alpha}(t)=\sum_{k\sigma}a_{\alpha k\sigma}^{\dagger}(t)a_{\alpha k\sigma}(t)$ is the particle number operator in the Heisenberg picture. In terms of the Keldysh Green's functions, the supercurrent can be formulated as
\begin{align}\label{current_Green}
\begin{split}
I_{\alpha}(t)=\frac{2e}{\hbar}\textrm{ReTr}\left\{\sigma_z\left[T_{\alpha}(t)G_{C\alpha}^<(t,t)\right]\right\},
\end{split}
\end{align}
where $\sigma_z$ is the Pauli matrix in Nambu space, $\textrm{Tr}$ denotes the trace over Nambu and spin spaces, and the lesser Green's function is defined as
\begin{align}\label{Keldysh_def}
\begin{split}
G_{C\alpha}^<(t_1,t_2)=i\sum_{k}\begin{pmatrix} \langle a_{\alpha k\uparrow}^{\dagger}(t_2)c_{\uparrow}(t_1)\rangle&\langle a_{\alpha k\downarrow}(t_2)c_{\uparrow}(t_1)\rangle\\ \langle a_{\alpha k\uparrow}^{\dagger}(t_2)c_{\downarrow}^{\dagger}(t_1)\rangle&\langle a_{\alpha k\downarrow}(t_2)c_{\downarrow}^{\dagger}(t_1)\rangle \end{pmatrix}.
\end{split}
\end{align}
The hopping matrix $T_{\alpha}(t)$ between the $\alpha$ lead and the central quantum dot is given by
\begin{align}\label{hoping_mat}
\begin{split}
T_{\alpha}(t)=\begin{pmatrix}t_{\alpha}(t)&0\\ 0&-t_{\alpha}^*(t)\end{pmatrix}.
\end{split}
\end{align}
Using Langreth's theorem, we can recast the supercurrent in Eq.~(\ref{current_Green}) as 
\begin{align}\label{current_int}
\begin{split}
I_{\alpha}(t)=\frac{2e}{\hbar}\int{dt_1}\textrm{ReTr}\left\{\sigma_z\left[G^r(t,t_1)\Sigma_{\alpha}^<(t_1,t)+G^<(t,t_1)\Sigma_{\alpha}^a(t_1,t)\right]\right\},
\end{split}
\end{align}
where $G^{r,<}(t,t_1)$ denote the retarded and lesser Green's functions of the central quantum dot, and $\Sigma_{\alpha}^{<,a}(t_1,t)$ are the corresponding self-energies arising from the coupling to the $\alpha$ lead. In the following, we consider two distinct regimes: (i) the purely dc case, where only a constant bias voltage is applied to the normal-metal top gate, and (ii) the ac case, where both a dc bias and an rf radiation are additionally applied to the system.

\section{I. Purely dc case}\label{sec_dc}
In the purely dc regime, the system reaches a non-equilibrium steady state. The current $I_\alpha$ becomes time-independent, and both the Green's functions and the self-energies depend solely on the time difference $t_1-t_2$. Consequently, we can perform a Fourier transform $D(t_1,t_2)=\int \frac{d\epsilon}{2\pi} e^{-i\epsilon(t_1-t_2)/\hbar} D(\epsilon)$ (where $D$ represents either a Green's function or a self-energy), which yields
 \begin{align}\label{current_int_energy}
\begin{split}
I_{\alpha}=\frac{2e}{h}\int{d\epsilon}\textrm{ReTr}\left\{\sigma_z\left[G^r(\epsilon)\Sigma_{\alpha}^<(\epsilon)+G^<(\epsilon)\Sigma_{\alpha}^a(\epsilon)\right]\right\}.
\end{split}
\end{align}
For the superconducting lead $\alpha$, the self-energies take the forms~\cite{Cuevas1996}
\begin{align}\label{SC_self_energy}
\begin{split}
&\Sigma_{\alpha}^{r,a}(\epsilon)=\frac{-i\pi t_{\alpha}^2\rho(\epsilon)}{\sqrt{\Delta^2-(\epsilon\pm i\gamma)^2}}\begin{pmatrix}-\epsilon\pm i\gamma&\Delta\exp{(-i\phi_{\alpha})}\\ \Delta\exp{(i\phi_{\alpha})}&-\epsilon\pm i\gamma\end{pmatrix},\\
&\Sigma_{\alpha}^{<}(\epsilon)=f(\epsilon)[\Sigma_{\alpha}^{a}(\epsilon)-\Sigma_{\alpha}^{r}(\epsilon)],\\
&\Sigma_{\alpha}^{>}(\epsilon)=[f(\epsilon)-1][\Sigma_{\alpha}^{a}(\epsilon)-\Sigma_{\alpha}^{r}(\epsilon)],
\end{split}
\end{align}
where $\gamma \to 0^+$ is an infinitesimal positive constant and $f(\epsilon)=1/[\exp{(\epsilon/k_BT)+1}]$ is the Fermi-Dirac distribution function at temperature $T$, with $k_B$ being the Boltzmann constant. In contrast, under the wide-band approximation, the self-energies of the normal-metal top gate take the form $\Sigma_{M}^{r,a}(\epsilon)=\mp i\pi t_M^2\rho(\epsilon)$ and $\Sigma_{M}^{<,>}(\epsilon)=[f_M(\epsilon,V)-0.5\pm0.5][\Sigma_{M}^{a}(\epsilon)-\Sigma_{M}^{r}(\epsilon)]$. Here, the Fermi distribution matrix for the biased metallic top gate is given by
\begin{align}\label{Fermi_metal}
\begin{split}
f_M(\epsilon,V)=\begin{pmatrix}1/\{\exp{[(\epsilon-eV)/k_BT]+1}\}&0\\0&1/\{\exp{[(\epsilon+eV)/k_BT]+1}\}\end{pmatrix}.
\end{split}
\end{align}

To calculate the supercurrent in Eq.~(\ref{current_int_energy}), we must first calculate the Green's functions $G^r(\epsilon)$ and $G^<(\epsilon)$ for the central quantum dot. They can be obtained by using the Dyson equation
\begin{align}\label{dc_Dyson}
\begin{split}
G^{r,a}(\epsilon)=[\epsilon\pm i\gamma-H_C-\sum_{\lambda}\Sigma_{\lambda}^{r,a}(\epsilon)]^{-1}
\end{split}
\end{align}
and the Keldysh equation
\begin{align}\label{dc_Keldysh}
\begin{split}
G^{<}(\epsilon)=G^{r}(\epsilon)\sum_{\lambda}\Sigma_{\lambda}^{<}(\epsilon)G^{a}(\epsilon).
\end{split}
\end{align}
Substituting Eqs.~(\ref{dc_Dyson}) and (\ref{dc_Keldysh}) into Eq.~(\ref{current_int_energy}) yields the supercurrent flowing from lead $\alpha$.

To proceed, we explore the supercurrent from each lead separately. For any lead $\alpha$, we rewrite the supercurrent in Eq.~(\ref{current_int_energy}) as 
 \begin{align}\label{current_int_energy_full}
\begin{split}
I_{\alpha}=\frac{e}{h}\int{d\epsilon}\textrm{Tr}\left\{\sigma_z\left[G^r(\epsilon)\Sigma_{\alpha}^<(\epsilon)+G^<(\epsilon)\Sigma_{\alpha}^a(\epsilon)-\Sigma_{\alpha}^<(\epsilon)G^a(\epsilon)-\Sigma_{\alpha}^r(\epsilon)G^<(\epsilon)\right]\right\}.
\end{split}
\end{align}
According to the Dyson equation $G^{r,a}(\epsilon)=[\epsilon\pm i\gamma-H_C-\sum_{\lambda}\Sigma_{\lambda}^{r,a}(\epsilon)]^{-1}$, we obtain 
\begin{align}\label{GS_relation}
\begin{split}
&G^r\Sigma_{\alpha}^<=G^r\Sigma_{\alpha}^<G^a(G^a)^{-1}=G^r\Sigma_{\alpha}^<G^a[\epsilon-H_C-\sum_{\lambda}\Sigma_{\lambda}^{a}(\epsilon)],\\
&\Sigma_{\alpha}^<G^a=(G^r)^{-1}G^r\Sigma_{\alpha}^<G^a=[\epsilon-H_C-\sum_{\lambda}\Sigma_{\lambda}^{r}(\epsilon)]G^r\Sigma_{\alpha}^<G^a.
\end{split}
\end{align}
Note that we have omitted the infinitesimal constant $\gamma$ for brevity, as it does not affect the final result. Therefore,
\begin{align}\label{G_m_relation}
\begin{split}
G^r\Sigma_{\alpha}^<-\Sigma_{\alpha}^<G^a&=G^r\Sigma_{\alpha}^<G^a(\epsilon-H_C-\sum_{\lambda}\Sigma_{\lambda}^{a})-(\epsilon-H_C-\sum_{\lambda}\Sigma_{\lambda}^{r})G^r\Sigma_{\alpha}^<G^a\\
&=\sum_{\lambda}\Sigma_{\lambda}^{r}G^r\Sigma_{\alpha}^<G^a-G^r\Sigma_{\alpha}^<G^a\sum_{\lambda}\Sigma_{\lambda}^{a}+[G^r\Sigma_{\alpha}^<G^a,\quad \epsilon-H_C],
\end{split}
\end{align}
where $[\cdots]$ denotes the commutator. Substituting Eq.~(\ref{G_m_relation}) into Eq.~(\ref{current_int_energy_full}), then using the cyclic property of the trace $\textrm{Tr}(\sigma_z[G^r\Sigma_{\alpha}^<G^a,\quad \epsilon-H_C])=\textrm{Tr}(G^r\Sigma_{\alpha}^<G^a[\epsilon-H_C,\quad \sigma_z])$ and the commutation relation $[\epsilon-H_C,\ \sigma_z]=0$, we arrive at
 \begin{align}\label{current_resolved}
\begin{split}
I_{\alpha}=\frac{e}{h}\int{d\epsilon}\textrm{Tr}\left\{\sigma_z\left[\sum_{\lambda}\Sigma_{\lambda}^{r}G^r\Sigma_{\alpha}^<G^a-G^r\Sigma_{\alpha}^<G^a\sum_{\lambda}\Sigma_{\lambda}^{a}+G^r\sum_{\lambda}\Sigma_{\lambda}^<G^a\Sigma_{\alpha}^a-\Sigma_{\alpha}^rG^r\sum_{\lambda}\Sigma_{\lambda}^<G^a\right]\right\},
\end{split}
\end{align}
where we have used the Keldysh equation defined in Eq.~(\ref{dc_Keldysh}) to calculate the other two terms in Eq.~(\ref{current_int_energy_full}). As a result, the supercurrent flowing from a superconducting lead $\alpha$ to lead $\beta$ assumes the equvilent form
 \begin{align}\label{current_resolved_form}
\begin{split}
I_{\alpha}^{\beta}&=\frac{e}{h}\int{d\epsilon}\textrm{Tr}\left[\sigma_z\left(\Sigma_{\beta}^{r}G^r\Sigma_{\alpha}^<G^a-G^r\Sigma_{\alpha}^<G^a\Sigma_{\beta}^{a}+G^r\Sigma_{\beta}^<G^a\Sigma_{\alpha}^a-\Sigma_{\alpha}^rG^r\Sigma_{\beta}^<G^a\right)\right]\\
&=\frac{2e}{h}\int{d\epsilon}\textrm{ReTr}\left[\sigma_z\left(\Sigma_{\beta}^{r}G^r\Sigma_{\alpha}^<G^a+G^r\Sigma_{\beta}^<G^a\Sigma_{\alpha}^a\right)\right],
\end{split}
\end{align}
where we have utilized the Hermitian conjugate relations $(\Sigma_{\beta}^{r}G^r\Sigma_{\alpha}^<G^a)^{\dagger}=-G^r\Sigma_{\alpha}^<G^a\Sigma_{\beta}^{a}$ and $(G^r\Sigma_{\beta}^<G^a\Sigma_{\alpha}^a)^{\dagger}=-\Sigma_{\alpha}^rG^r\Sigma_{\beta}^<G^a$ in deriving the final expression. Equation~(\ref{current_resolved_form}) represents the general formula for the terminal-resolved supercurrent flowing from lead $\alpha$ to lead $\beta$. It is evident that these resolved supercurrents scale as $t_{\alpha}^2t_{\beta}^2$ via the self-energies, and strictly satisfy the current conservation relation $I_{\alpha}^{\beta}=-I_{\beta}^{\alpha}$.

Alternatively, because the self-energy induced by the coupling to the normal-metal top gate is diagonal in Nambu space, the current from this metallic lead can be evaluated as
 \begin{align}\label{current_metal}
\begin{split}
I_{M}&=\frac{e}{h}\int{d\epsilon}\textrm{Tr}\left\{\sigma_z\left[G^r\Sigma_{M}^<+G^<\Sigma_{M}^a-\Sigma_{M}^<G^a+\Sigma_{M}^rG^<\right]\right\}\\
&=\frac{e}{h}\int{d\epsilon}\textrm{Tr}\left\{\sigma_z\left[G^r\Sigma_{M}^<-G^a\Sigma_{M}^<+G^<\Sigma_{M}^a-G^<\Sigma_{M}^r\right]\right\}\\
&=\frac{e}{h}\int{d\epsilon}\textrm{Tr}\left\{\sigma_z\left[(G^r-G^a)\Sigma_{M}^<+G^<(\Sigma_{M}^a-\Sigma_{M}^r)\right]\right\}\\
&=\frac{e}{h}\int{d\epsilon}\textrm{Tr}\left\{\sigma_z\left[(G^>-G^<)\Sigma_{M}^<+G^<(\Sigma_{M}^<-\Sigma_{M}^>)\right]\right\}\\
&=\frac{e}{h}\int{d\epsilon}\textrm{Tr}\left[\sigma_z\left(G^>\Sigma_{M}^<-G^<\Sigma_{M}^>\right)\right],
\end{split}
\end{align}
where we have applied the identities $G^r-G^a=G^>-G^<$ and $\Sigma_{M}^a-\Sigma_{M}^r=\Sigma_{M}^<-\Sigma_{M}^>$. Substituting the full Keldysh equation [Eq.~(\ref{dc_Keldysh})] into Eq.~(\ref{current_metal}) yields
\begin{align}\label{dc_current_metal}
\begin{split}
I_{M}=\frac{e}{h}\int{d\epsilon}\textrm{Tr}\left[\sigma_z\left(G^r\sum_{\lambda}\Sigma_{\lambda}^>G^a\Sigma_{M}^<-G^r\sum_{\lambda}\Sigma_{\lambda}^<G^a\Sigma_{M}^>\right)\right], 
\end{split}
\end{align}
and the resolved current from the metallic lead is correspondingly given by
\begin{align}\label{dc_current_metal_resolved}
\begin{split}
I_{M}^{\alpha}=\frac{e}{h}\int{d\epsilon}\textrm{Tr}\left[\sigma_z\left(G^r\Sigma_{\alpha}^>G^a\Sigma_{M}^<-G^r\Sigma_{\alpha}^<G^a\Sigma_{M}^>\right)\right].
\end{split}
\end{align}
For $\alpha = \text{L, R}$, this term represents the quasiparticle tunneling current from the normal-metal lead to the superconducting lead, which is exponentially suppressed in the subgap regime ($eV \ll \Delta_0$) due to the large superconducting gap. Consequently, the low-bias current flowing from the metallic lead M is overwhelmingly dominated by the Andreev reflection processes, evaluated as
\begin{align}\label{dc_current_metal_Andreev}
\begin{split}
I_{M}^{M}&=\frac{e}{h}\int{d\epsilon}\textrm{Tr}\left[\sigma_z\left(G^r\Sigma_{M}^>G^a\Sigma_{M}^<-G^r\Sigma_{M}^<G^a\Sigma_{M}^>\right)\right]\\
&\approx \frac{2e^2V}{h}\textrm{Tr}[G_{\uparrow\downarrow}^r\Gamma_{\downarrow\downarrow}^MG_{\downarrow\uparrow}^a\Gamma_{\uparrow\uparrow}^M-G_{\downarrow\uparrow}^r\Gamma_{\uparrow\uparrow}^MG_{\uparrow\downarrow}^a\Gamma_{\downarrow\downarrow}^M],
\end{split}
\end{align}
where $\Gamma^M_{\sigma\sigma}(\epsilon)=i[\Sigma_{M;\sigma\sigma}^r-\Sigma_{M;\sigma\sigma}^a]$ is the linewidth function of the metallic lead. It is physically intuitive that this $I_M^M$ is proportional to the applied top-gate bias $V$ and scales with $t_M^4$ via the self-energies. Furthermore, imposing the steady-state charge conservation condition yields the relations between the terminal-resolved currents:
\begin{align}\label{current_relation}
\begin{split}
I_L^R=-I_R^L,\quad I_L^M+I_R^M=-I_M^M.
\end{split}
\end{align}
Moreover, as implied by Eq.~(\ref{current_resolved_form}), because $I_L^M$ and $I_R^M$ scale proportionally with $t_L^2t_M^2$ and $t_R^2t_M^2$, respectively, they can be re-expressed in terms of the total Andreev current from the metallic lead as $I_L^M=-t_L^2 I_M^M/(t_L^2+t_R^2)$ and $I_R^M=-t_R^2 I_M^M/(t_L^2+t_R^2)$. Consequently, the net supercurrent flowing across the left and right superconducting leads consists of two distinct components:
\begin{align}\label{supercurrent}
\begin{split}
I_s=(I_{L}^R-I_R^L)/2+(I_{L}^M-I_R^M)/2=(I_{L}^R-I_R^L)/2+\frac{(t_R^2-t_L^2)I_M^M}{2(t_L^2+t_R^2)}.
\end{split}
\end{align}

\section{II. ac case}\label{sec_ac}
We next turn to the ac case, where a dc bias $V_{dc}$ and an rf radiation field, characterized by $V_{ac}\cos{\omega t}$, are applied to the system alongside the top-gate bias. Here, $V_{ac}$ and $\omega$ denote the amplitude and frequency of the rf radiation, respectively. In this non-equilibrium regime, the on-site energy for each terminal acquires a time-dependent modulation $\epsilon_{\alpha}(t)=\epsilon_{\alpha}-eV_{\alpha}^{dc}-eV_{\alpha}^{ac}\cos{\omega t}$, where $V_{\alpha}^{dc}$ and $V_{\alpha}^{ac}$ are the respective dc and ac voltage drops at lead $\alpha$. Concurrently, the superconducting order parameter dynamically evolves its phase as $\Delta_{\alpha}(t)=\Delta_0\exp{[-i\phi_{\alpha}-(i2e/\hbar)\int_0^t (V_{\alpha}^{dc}+V_{\alpha}^{ac}\cos{\omega\tau}) d\tau]}$. By employing a gauge transformation $U(t)=\exp{\{\sum_{\alpha k\sigma}(i/\hbar)[(\phi_{\alpha}/2)+2e\int_0^t (V_{\alpha}^{dc}+V_{\alpha}^{ac}\cos{\omega\tau}) d\tau]a_{\alpha k\sigma}^{\dagger}a_{\alpha k\sigma}\}}$~\cite{Sun2000}, we can completely eliminate the explicit time dependence in the lead and central-region Hamiltonians, transferring it entirely to the tunneling Hamiltonian $H_T(t)$ as a time-dependent Peierls phase factor: $H_T(t)=\sum_{\alpha k\sigma}[\tilde{t}_{\alpha}(t)a_{\alpha k\sigma}^{\dagger}c_{\sigma}+\textrm{h.c.}]$, where $\tilde{t}_{\alpha}(t)=t_{\alpha}\exp{\{(-i/2)[\phi_{\alpha}+2eV_{dc}^{\alpha}t/\hbar+(2eV_{ac}^{\alpha}/\hbar\omega)\sin{\omega t}]\}}$. Accordingly, the hopping matrix in Eq.~(\ref{hoping_mat}) becomes
\begin{align}\label{hoping_mat_time}
\begin{split}
T_{\alpha}(t)=\begin{pmatrix}t_{\alpha}\exp{\{(-i/2)[\phi_{\alpha}+2eV_{dc}^{\alpha}t/\hbar+(2eV_{ac}^{\alpha}/\hbar\omega)\sin{\omega t}]\}}&0\\ 0&-t_{\alpha}^*\exp{\{(i/2)[\phi_{\alpha}+2eV_{dc}^{\alpha}t/\hbar+(2eV_{ac}^{\alpha}/\hbar\omega)\sin{\omega t}]\}}\end{pmatrix}.
\end{split}
\end{align}
Because the Green's functions and self-energies in Eq.~(\ref{current_int}) are no longer solely functions of the time difference, we must perform a double Fourier transform with respect to $t_1$ and $t_2$ independently. The double Fourier transform is defined as
\begin{align}\label{double_Fourier}
\begin{split}
&D(t_1,t_2)=\frac{1}{2\pi}\int{d\epsilon_1d\epsilon_2e^{-i\epsilon_1t_1/\hbar}e^{i\epsilon_2t_2/\hbar}D(\epsilon_1,\epsilon_2)},\\
&D(\epsilon_1,\epsilon_2)=\frac{1}{2\pi}\int{dt_1dt_2e^{i\epsilon_1t_1/\hbar}e^{-i\epsilon_2t_2/\hbar}D(t_1,t_2)}.
\end{split}
\end{align}
Substituting this double Fourier transform into Eq.~(\ref{current_int}) yields
\begin{align}\label{ac_current_int}
\begin{split}
I_{\alpha}(t)=\frac{2e}{h}\int{d\epsilon d\epsilon_1 d\epsilon^{\prime}e^{-i(\epsilon-\epsilon^{\prime})t/\hbar}\textrm{ReTr}\left\{\sigma_z\left[G^r(\epsilon,\epsilon_1)\Sigma_{\alpha}^<(\epsilon_1,\epsilon^{\prime})+G^<(\epsilon,\epsilon_1)\Sigma_{\alpha}^a(\epsilon_1,\epsilon^{\prime})\right]\right\} }.
\end{split}
\end{align}

We next evaluate the retarded self-energy, which takes the form
 \begin{align}\label{retarted_self_time}
\begin{split}
\Sigma_{\alpha}^r(t_1,t_2)=T_{\alpha}(t_1)g^r_{\alpha}(t_1,t_2)T_{\alpha}^*(t_2)=\begin{pmatrix}\Sigma_{\alpha;\uparrow\uparrow}^r(t_1,t_2)&\Sigma_{\alpha;\uparrow\downarrow}^r(t_1,t_2)\\ \Sigma_{\alpha;\downarrow\uparrow}^r(t_1,t_2)&\Sigma_{\alpha;\downarrow\downarrow}^r(t_1,t_2)\end{pmatrix}.
\end{split}
\end{align}
Here, $\Sigma_{\alpha;\sigma_1\sigma_2}^r(t_1,t_2)=T_{\alpha;\sigma_1}(t_1)g^r_{\alpha;\sigma_1\sigma_2}(t_1,t_2)T_{\alpha;\sigma_2}^*(t_2)$, where $g^r_{\alpha;\sigma_1\sigma_2}(t_1,t_2)$ is the free-electron Green's function of the lead. Utilizing the Jacobi-Anger expansion $\exp{\left(iz\sin{x}\right)}=\sum_{n=-\infty}^{\infty}\mathcal{J}_n(z)\exp{\left(inx\right)}$, where $\mathcal{J}_n(z)$ is the $n$-th order Bessel function of the first kind, we can express the components of the self-energy as 
\begin{align}\label{self_energy_t}
\begin{split}
&\Sigma_{\alpha;\uparrow\uparrow}^r(t_1,t_2)=t_{\alpha}^2\sum_{m,n}\mathcal{J}_m(-\kappa_\alpha)\mathcal{J}_n(\kappa_\alpha)\exp{\left\{i\left[\left(eV_{dc}^{\alpha}/\hbar+n\omega\right)t_2-\left(eV_{dc}^{\alpha}/\hbar+m\omega\right)t_1\right]\right\}}g_{\alpha;\uparrow\uparrow}^r(t_1,t_2),\\
&\Sigma_{\alpha;\uparrow\downarrow}^r(t_1,t_2)=-t_{\alpha}^2\sum_{m,n}\mathcal{J}_m(-\kappa_\alpha)\mathcal{J}_n(-\kappa_\alpha)\exp{\left\{-i\left[\phi_{\alpha}+\left(eV_{dc}^{\alpha}/\hbar+n\omega\right)t_2+\left(eV_{dc}^{\alpha}/\hbar+m\omega\right)t_1\right]\right\}}g_{\alpha;\uparrow\downarrow}^r(t_1,t_2),\\
&\Sigma_{\alpha;\downarrow\uparrow}^r(t_1,t_2)=-t_{\alpha}^2\sum_{m,n}\mathcal{J}_m(\kappa_\alpha)\mathcal{J}_n(\kappa_\alpha)\exp{\left\{i\left[\phi_{\alpha}+\left(eV_{dc}^{\alpha}/\hbar+n\omega\right)t_2+\left(eV_{dc}^{\alpha}/\hbar+m\omega\right)t_1\right]\right\}}g_{\alpha;\downarrow\uparrow}^r(t_1,t_2),\\
&\Sigma_{\alpha;\downarrow\downarrow}^r(t_1,t_2)=t_{\alpha}^2\sum_{m,n}\mathcal{J}_m(\kappa_\alpha)\mathcal{J}_n(-\kappa_\alpha)\exp{\left\{-i\left[\phi_{\alpha}+\left(eV_{dc}^{\alpha}/\hbar+n\omega\right)t_2-\left(eV_{dc}^{\alpha}/\hbar+m\omega\right)t_1\right]\right\}}g_{\alpha;\downarrow\downarrow}^r(t_1,t_2),
\end{split}
\end{align}
where $\kappa_{\alpha}=eV_{ac}^{\alpha}/\hbar\omega$ is the scaled ac amplitude. Imposing the double Fourier transform recasts these matrix elements into the energy domain as
\begin{align}\label{self_energy_E}
\begin{split}
&\Sigma_{\alpha;\uparrow\uparrow}^r(\epsilon_1,\epsilon_2)=t_{\alpha}^2\sum_{m,n}\mathcal{J}_m(-\kappa_\alpha)\mathcal{J}_n(\kappa_\alpha)g_{\alpha;\uparrow\uparrow}^r(\epsilon_1-eV_{dc}^{\alpha}-m\hbar\omega)\delta(\epsilon_2-\epsilon_1+(m-n)\hbar\omega),\\
&\Sigma_{\alpha;\uparrow\downarrow}^r(\epsilon_1,\epsilon_2)=-t_{\alpha}^2\exp{\left(-i\phi_{\alpha}\right)}\sum_{m,n}\mathcal{J}_m(-\kappa_\alpha)\mathcal{J}_n(-\kappa_\alpha)g_{\alpha;\uparrow\downarrow}^r(\epsilon_1-eV_{dc}^{\alpha}-m\hbar\omega)\delta(\epsilon_2-\epsilon_1+2eV_{dc}^{\alpha}+(m+n)\hbar\omega),\\
&\Sigma_{\alpha;\downarrow\uparrow}^r(\epsilon_1,\epsilon_2)=-t_{\alpha}^2\exp{\left(i\phi_{\alpha}\right)}\sum_{m,n}\mathcal{J}_m(\kappa_\alpha)\mathcal{J}_n(\kappa_\alpha)g_{\alpha;\downarrow\uparrow}^r(\epsilon_1+eV_{dc}^{\alpha}+m\hbar\omega)\delta(\epsilon_2-\epsilon_1-2eV_{dc}^{\alpha}-(m+n)\hbar\omega),\\
&\Sigma_{\alpha;\downarrow\downarrow}^r(\epsilon_1,\epsilon_2)=t_{\alpha}^2\sum_{m,n}\mathcal{J}_m(\kappa_\alpha)\mathcal{J}_n(-\kappa_\alpha)g_{\alpha;\downarrow\downarrow}^r(\epsilon_1+eV_{dc}^{\alpha}+m\hbar\omega)\delta(\epsilon_2-\epsilon_1-(m-n)\hbar\omega),
\end{split}
\end{align}
where $\delta(x)$ is the Dirac delta function. Equation~(\ref{self_energy_E}) explicitly shows that the self-energy $\Sigma_{\alpha}^r(\epsilon_1,\epsilon_2)$ is non-vanishing only when the resonance condition $\epsilon_2-\epsilon_1-2meV_{dc}^{\alpha}-n\hbar\omega=0$ is satisfied for integers $m$ and $n$. For notational convenience, we introduce the Floquet-like component representation $D_{pq}^{mn}(\epsilon)=D(\epsilon+peV_{dc}+m\hbar\omega,\epsilon+qeV_{dc}+n\hbar\omega)$. With this notation, Eq.~(\ref{self_energy_E}) can be compactly expressed as
\begin{align}\label{self_energy_E_mnpq}
\begin{split}
&\Sigma_{\alpha;\uparrow\uparrow;pq}^{r;mn}(\epsilon)=t_{\alpha}^2\sum_{k}\mathcal{J}_{m-k}(-\kappa_\alpha)\mathcal{J}_{n-k}(\kappa_\alpha)g_{\alpha;\uparrow\uparrow}^r(\epsilon+peV_{dc}-eV_{dc}^{\alpha}+k\hbar\omega)\delta(p,q),\\
&\Sigma_{\alpha;\uparrow\downarrow;pq}^{r;mn}(\epsilon)=-t_{\alpha}^2\exp{\left(-i\phi_{\alpha}\right)}\sum_{k}\mathcal{J}_{k-m}(-\kappa_\alpha)\mathcal{J}_{n-k}(-\kappa_\alpha)g_{\alpha;\uparrow\downarrow}^r(\epsilon+peV_{dc}-eV_{dc}^{\alpha}+k\hbar\omega)\delta((q-p)eV_{dc}+2eV_{dc}^\alpha),\\
&\Sigma_{\alpha;\downarrow\uparrow;pq}^{r;mn}(\epsilon)=-t_{\alpha}^2\exp{\left(i\phi_{\alpha}\right)}\sum_{k}\mathcal{J}_{k-m}(\kappa_\alpha)\mathcal{J}_{k-n}(\kappa_\alpha)g_{\alpha;\downarrow\uparrow}^r(\epsilon+peV_{dc}+eV_{dc}^{\alpha}+k\hbar\omega)\delta((q-p)eV_{dc}-2eV_{dc}^\alpha),\\
&\Sigma_{\alpha;\downarrow\downarrow;pq}^{r;mn}(\epsilon)=t_{\alpha}^2\sum_{k}\mathcal{J}_m(\kappa_\alpha)\mathcal{J}_n(-\kappa_\alpha)g_{\alpha;\downarrow\downarrow}^r(\epsilon+peV_{dc}+eV_{dc}^{\alpha}+k\hbar\omega)\delta(p,q).
\end{split}
\end{align}
Similarly, the advanced and lesser self-energies have the corresponding forms
\begin{align}\label{dis_self_energy_E_mnpq}
\begin{split}
&\Sigma_{\alpha;\uparrow\uparrow;pq}^{a,<;mn}(\epsilon)=t_{\alpha}^2\sum_{k}\mathcal{J}_{m-k}(-\kappa_\alpha)\mathcal{J}_{n-k}(\kappa_\alpha)g_{\alpha;\uparrow\uparrow}^{a,<}(\epsilon+peV_{dc}-eV_{dc}^{\alpha}+k\hbar\omega)\delta(p,q),\\
&\Sigma_{\alpha;\uparrow\downarrow;pq}^{a,<;mn}(\epsilon)=-t_{\alpha}^2\exp{\left(-i\phi_{\alpha}\right)}\sum_{k}\mathcal{J}_{k-m}(-\kappa_\alpha)\mathcal{J}_{n-k}(-\kappa_\alpha)g_{\alpha;\uparrow\downarrow}^{a,<}(\epsilon+peV_{dc}-eV_{dc}^{\alpha}+k\hbar\omega)\delta((q-p)eV_{dc}+2eV_{dc}^\alpha),\\
&\Sigma_{\alpha;\downarrow\uparrow;pq}^{a,<;mn}(\epsilon)=-t_{\alpha}^2\exp{\left(i\phi_{\alpha}\right)}\sum_{k}\mathcal{J}_{k-m}(\kappa_\alpha)\mathcal{J}_{k-n}(\kappa_\alpha)g_{\alpha;\downarrow\uparrow}^{a,<}(\epsilon+peV_{dc}+eV_{dc}^{\alpha}+k\hbar\omega)\delta((q-p)eV_{dc}-2eV_{dc}^\alpha),\\
&\Sigma_{\alpha;\downarrow\downarrow;pq}^{a,<;mn}(\epsilon)=t_{\alpha}^2\sum_{k}\mathcal{J}_m(\kappa_\alpha)\mathcal{J}_n(-\kappa_\alpha)g_{\alpha;\downarrow\downarrow}^{a,<}(\epsilon+peV_{dc}+eV_{dc}^{\alpha}+k\hbar\omega)\delta(p,q).
\end{split}
\end{align}
Furthermore, the Green's functions are given by the Dyson equation
\begin{align}\label{ac_Dyson}
\begin{split}
&G^{r;mn}_{C;pq}(\epsilon)=g_{C;p}^{r;m}(\epsilon)\delta_{mn}\delta_{pq}+\sum_{\alpha,k,l}g_{C;p}^{r;m}(\epsilon)\Sigma^{r;mk}_{\alpha;pl}(\epsilon)G^{r;kn}_{C;lq}(\epsilon),\\
&G^{a;mn}_{C;pq}(\epsilon)=g_{C;p}^{a;m}(\epsilon)\delta_{mn}\delta_{pq}+\sum_{\alpha,k,l}g_{C;p}^{a;m}(\epsilon)\Sigma^{a;mk}_{\alpha;pl}(\epsilon)G^{a;kn}_{C;lq}(\epsilon),
\end{split}
\end{align}
and the Keldysh equation
\begin{align}\label{ac_Keldysh}
\begin{split}
&G^{<;mn}_{C;pq}(\epsilon)=\sum_{\alpha,k,l,i,j}G_{C;pl}^{r;mk}(\epsilon)\Sigma^{<;ki}_{\alpha;lj}(\epsilon)G^{a;in}_{C;jq}(\epsilon),\\
&G^{>;mn}_{C;pq}(\epsilon)=\sum_{\alpha,k,l,i,j}G_{C;pl}^{r;mk}(\epsilon)\Sigma^{>;ki}_{\alpha;lj}(\epsilon)G^{a;in}_{C;jq}(\epsilon),
\end{split}
\end{align}
where $g_{C;p}^{r,a;m}(\epsilon)=(\epsilon\pm i\gamma+peV_{dc}+m\hbar\omega-H_C)^{-1}$. Equations~(\ref{ac_Dyson}) and (\ref{ac_Keldysh}) can be solved self-consistently by numerically truncating the Floquet sub/superscript indices at a sufficiently large cutoff~\cite{Li2018}. Finally, the time-dependent supercurrent in Eq.~(\ref{ac_current_int}) can be reconstructed as
\begin{align}\label{ac_current_sum}
\begin{split}
I_{\alpha}(t)=\sum_{p,m}\exp{[i(peV_{dc}/\hbar+m\omega)t]}I_{\alpha;p}^m,
\end{split}
\end{align}
with the supercurrent components
\begin{align}\label{ac_current_resolved}
\begin{split}
I_{\alpha;p}^m=\frac{2e}{h}\sum_{q,n}\int{d\epsilon \textrm{ReTr}\left\{\sigma_z\left[G_{C;0q}^{r;0n}(\epsilon)\Sigma_{\alpha;qp}^{<;nm}(\epsilon)+G_{C;0q}^{<;0n}(\epsilon)\Sigma_{\alpha;qp}^{a;nm}(\epsilon)\right]\right\} }.
\end{split}
\end{align}
The net dc current constitutes three distinct components: $I_\alpha = I_{\alpha;0}^0(0) + I_{\alpha;0}^0(V_{dc}) + \sum_{p,m}I_{\alpha;p}^m \delta_{2peV_{dc}, -m\hbar\omega}$. Here, the first term $I_{\alpha;0}^0(0)=I_s(\phi)$ corresponds to the purely phase-dependent supercurrent under rf irradiation (the 0th-order Shapiro step), the second term $I_{\alpha;0}^0(V_{dc})$ represents the dissipative background quasiparticle current, and the third term encapsulates the higher-order Shapiro steps rigidly localized at the phase-locked resonance conditions.

Specifically, when the system is driven solely by the rf radiation without a dc bias, the phase-dependent supercurrent $I_{\alpha;0}^0(0)=I_s(\phi)$ must be evaluated independently. In this limit, the system is modulated by a single frequency $\omega$ with the sub-index $pd$ eliminated completely, and the Floquet self-energies simplify to
\begin{align}\label{self_energy_E_mn}
\begin{split}
&\Sigma_{\alpha;\uparrow\uparrow}^{r;mn}(\epsilon)=t_{\alpha}^2\sum_{k}\mathcal{J}_{m-k}(-\kappa_\alpha)\mathcal{J}_{n-k}(\kappa_\alpha)g_{\alpha;\uparrow\uparrow}^r(\epsilon+k\hbar\omega),\\
&\Sigma_{\alpha;\uparrow\downarrow}^{r;mn}(\epsilon)=-t_{\alpha}^2\exp{\left(-i\phi_{\alpha}\right)}\sum_{k}\mathcal{J}_{k-m}(-\kappa_\alpha)\mathcal{J}_{n-k}(-\kappa_\alpha)g_{\alpha;\uparrow\downarrow}^r(\epsilon+k\hbar\omega),\\
&\Sigma_{\alpha;\downarrow\uparrow}^{r;mn}(\epsilon)=-t_{\alpha}^2\exp{\left(i\phi_{\alpha}\right)}\sum_{k}\mathcal{J}_{k-m}(\kappa_\alpha)\mathcal{J}_{k-n}(\kappa_\alpha)g_{\alpha;\downarrow\uparrow}^r(\epsilon+k\hbar\omega),\\
&\Sigma_{\alpha;\downarrow\downarrow}^{r;mn}(\epsilon)=t_{\alpha}^2\sum_{k}\mathcal{J}_m(\kappa_\alpha)\mathcal{J}_n(-\kappa_\alpha)g_{\alpha;\downarrow\downarrow}^r(\epsilon+k\hbar\omega),
\end{split}
\end{align}
and
\begin{align}\label{dis_self_energy_E_mn}
\begin{split}
&\Sigma_{\alpha;\uparrow\uparrow}^{a,<;mn}(\epsilon)=t_{\alpha}^2\sum_{k}\mathcal{J}_{m-k}(-\kappa_\alpha)\mathcal{J}_{n-k}(\kappa_\alpha)g_{\alpha;\uparrow\uparrow}^{a,<}(\epsilon+k\hbar\omega),\\
&\Sigma_{\alpha;\uparrow\downarrow}^{a,<;mn}(\epsilon)=-t_{\alpha}^2\exp{\left(-i\phi_{\alpha}\right)}\sum_{k}\mathcal{J}_{k-m}(-\kappa_\alpha)\mathcal{J}_{n-k}(-\kappa_\alpha)g_{\alpha;\uparrow\downarrow}^{a,<}(\epsilon+k\hbar\omega),\\
&\Sigma_{\alpha;\downarrow\uparrow}^{a,<;mn}(\epsilon)=-t_{\alpha}^2\exp{\left(i\phi_{\alpha}\right)}\sum_{k}\mathcal{J}_{k-m}(\kappa_\alpha)\mathcal{J}_{k-n}(\kappa_\alpha)g_{\alpha;\downarrow\uparrow}^{a,<}(\epsilon+k\hbar\omega),\\
&\Sigma_{\alpha;\downarrow\downarrow}^{a,<;mn}(\epsilon)=t_{\alpha}^2\sum_{k}\mathcal{J}_m(\kappa_\alpha)\mathcal{J}_n(-\kappa_\alpha)g_{\alpha;\downarrow\downarrow}^{a,<}(\epsilon+k\hbar\omega).
\end{split}
\end{align}
 Substituting these self-energies into the Dyson equation and the Keldysh equation, we can obtain the Green's function, which further yields the supercurrent $I_{\alpha;0}^0(0)=I_s(\phi)$ by using Eq.~\ref{ac_current_resolved}.